\documentclass[12pt]{article}
\usepackage{amssymb,amsmath}
\usepackage[noblocks]{authblk}
\usepackage[top=0.75in, bottom=0.75in, left=0.75in, right=0.75in, dvips]{geometry}
\usepackage{caption}  
\date{}   

\begin{document}
\textwidth 10.0in 
\textheight 9.0in 
\topmargin -0.60in
\title{A Canonical Analysis of the Massless Superparticle}
\author[1,2]{D.G.C. McKeon\thanks{Email: dgmckeo2@uwo.ca}}
\affil[1] {Department of Applied Mathematics, The
University of Western Ontario, London, ON N6A 5B7, Canada}
\affil[2] {Department of Mathematics and
Computer Science, Algoma University, Sault St.Marie, ON P6A
2G4, Canada}
\maketitle

\maketitle

\begin{abstract}
The canonical structure of the action for a massless superparticle is considered in $d = 2+1$ and $d = 3+1$ dimensions.  This is done by examining the contribution to the action of each of the components of the spinor $\theta$ present; no attempt is made to maintain manifest covariance.  Upon using the Dirac Bracket to eliminate the second class constraints arising from the canonical momenta associated with half of these components, we find that the remaining components have canonical momenta that are all first class constraints.  From these first class constraints, it is possible to derive the generator of half of the local Fermionic $\kappa$-symmetry of Siegel; which half is contingent upon the choice of which half of the momenta associated with the components of $\theta$ are taken to be second class constraints.  The algebra of the generator of this Fermionic symmetry transformation is examined.
\end{abstract}

\section{Introduction}

There have been two extensions of the relativistic Lagrangian for a particle that are supersymmetric.  The spinning particle is locally supersymmetric on the world line of the particle's trajectory [1,2]; the superparticle is globally supersymmetric in its target space [3,4].  It was soon discovered [5] that there is also a local supersymmetry transformation that leaves the superparticle action invariant.  This so called $\kappa$-symmetry has made quantization of the superparticle quite difficult [6-10, 20, 21].

Dirac pointed out that the presence of first class constraints in a theory is connected with a gauge invariance of the action [17, 18].  The construction of the generator of such a gauge invariance from these first class constraints has been accomplished in two ways.  One way is to examine the invariance of the total action in phase space [11] (the HTZ approach); the other is to consider local transformations that leave the equations of motion invariant [12] (the $C$ approach).  Initially these procedures were just applied to Bosonic gauge symmetries; indeed it was possible to uncover a gauge symmetry in the Einstein-Hilbert action in $d = 1 + 1$ dimensions that is distinct from the usual diffeomorphism invariance by using these approaches [13].  It has also proved possible to make use of these ways of deriving the generator of a gauge transformation to find Fermionic (super) gauge transformations in such models as the spinning particle [14], supergravity in $(2 + 1)d$ [15] and a model involving a non-Abelian vector coupled to a Majorana spin 3/2 field in $(3 + 1)d$ [16].  Below, we shall analyse the superparticle action using the Dirac procedure and show how the first class constraints in this theory are related to $\kappa$-symmetry.  Conventions used appear in the appendix. 

\section{The Superparticle}

We shall consider the action [1,2]
\begin{equation}
S = \int d\tau \frac{1}{2e(\tau)} Y^\mu(\tau)Y_\mu(\tau)
\end{equation}
where
\begin{equation}
Y^\mu(\tau) = \dot{x}^\mu(\tau) + i \dot{\overline{\theta}}(\tau) \gamma^\mu\theta(\tau).
\end{equation}
The field $e$ is a ``Lagrange multiplier'' field, $x^\mu(\tau)$ is a Bosonic vector in $d$ dimensions and $\theta$ is a Fermionic (Grassmann) spinor.  We restrict our attention to $d = 2+1$ and $d = 3+1$ and so it is possible to take $\theta$ to be Majorana.

In performing a canonical analysis of $S$ [17, 18], we first find that the momenta conjugate to $e$, $x^\mu$ and $\overline{\theta}$ to be 
\begin{subequations}
\begin{align}
p_e &= 0 \\
p_\mu &= \frac{1}{e} (\dot{x}_\mu + i \dot{\overline{\theta}} \gamma_\mu \theta) = \frac{1}{e}\eta_{\mu\nu} Y^\nu \\
\pi &= \frac{i}{e} (\gamma_\mu\theta) (\dot{x}^\mu + i \dot{\overline{\theta}} \gamma^\mu \theta) = ip\!\!\!/ \theta .
\end{align}
\end{subequations}
From these equations and eq. (A.8) it follows that the canonical momentum is
\begin{equation}
H_c = \frac{e}{2} p^2. 
\end{equation}
Eqs. (3a,c) evidently are a set of primary (first generation) constraints. Together, eqs. (3a) and (4) lead to the second generation constraint
\begin{equation}
p^2 = 0.
\end{equation}
The status of the constraint of eq. (3c) is not immediately clear as 
$p\!\!\!/ \pi = ip^2\theta$, and by eq. (5), this weakly vanishes.  If one were to simply treat
\begin{equation}
\sigma = \pi - i p\!\!\!/ \theta
\end{equation}
as a set of second class constraints satisfying the Poisson Bracket (PB) algebra
\begin{equation}
\left\lbrace \sigma_i, \overline{\sigma}_j \right\rbrace = - 2i p\!\!\!/_{ij}
\end{equation}
(In eq. (7) we have $\overline{\sigma} = \overline{\pi} + i \overline{\theta}p\!\!\!/$ and as by eq. (A.10) $\left\lbrace \pi_i, \overline{\theta}_j \right\rbrace = -\delta_{ij}$ we see from the Majorana condition that $\left\lbrace \theta_i, \overline{\pi}_j \right\rbrace = + \delta_{ij}$; eq. (7) now follows.) The Dirac Bracket (DB) that follows from eq. (7) is
\begin{equation}
\left\lbrace A, B \right\rbrace^* = \left\lbrace A, B \right\rbrace - 
\left\lbrace A, \overline{\sigma}_i \right\rbrace \left(\frac{i  p\!\!\!/}{2p^2}\right)_{ij} \left\lbrace \sigma_j, B \right\rbrace .
\end{equation}
This ensures that $\left\lbrace A, \overline{\sigma}_k\right\rbrace^* = 0$. (With Fermionic second class constraints one can have an odd number of constraints.)  The occurrence of $1/p^2$ in eq. (8) indicates that the DB is ill defined when the first class constraint of eq. (5) is applied; this is a situation that was not was not directly treated by Dirac in his discussion of constrained systems.  It is apparent that since $ p\!\!\!/ \pi$ weakly vanishes then some of the components of $\sigma$ are first class, others are second class.  Disentangling these two types of constraints in a manifestly covariant fashion has been attempted in a number of ways [5-10].

In any case, if one were to adopt the DB of eq. (8) then are encounters such unusual results as 
\begin{equation}
\left\lbrace x^\mu, x^\nu \right\rbrace^* = - \frac{i\overline{\theta}\gamma^\mu  p\!\!\!/ \gamma^\nu \theta}{p^2} ;
\end{equation}
such relations also appear in ref. [2].

Rather than trying to retain manifest covariance when performing a canonical analysis of the action of eq. (1), we shall examine this action when it is expressed in terms of the components of the spinor $\theta$.  This not only leads to a loss of general covariance, but also necessitates having to deal with each dimension $d$ of the target space individually.  However there are two advantages of this way of analysing the action.  First of all, it becomes immediately apparent how $\sigma$ is to be decomposed into these components that are first class constraints and those  that are second class constraints.  Secondly, it becomes possible to relate the first class constraints to a generator $G$ that gives rise to Fermionic gauge transformations related to the $\kappa$-symmetry of Siegel [5]. This $\kappa$-symmetry is given by
\[
 \delta x^\mu = i\overline{\theta} \gamma^\mu \delta\theta, \quad \delta\theta = i\gamma \cdot p\kappa, \quad \delta e = 4e\dot{\overline{\theta}}\kappa \eqno(10a-c)
\]
where $\kappa(\tau)$ is a local Fermionic gauge function.

We will now examine in turn the cases $d = 3$ and $d = 4$.  Conventions appear in the appendix.

\section{d = 2+1}

When $\theta$ is a Majorana spinor in $d = 3$ dimensions it can be written as 
\[ \theta = \left( \begin{array}{c}
u \\
d\end{array}
\right) \eqno(11) \]
where $u$ and $d$ are real Fermionic variables. The action of eq. (1) then becomes
\[
S = \int d\tau \frac{1}{2e(\tau)} \bigg[ (\dot{x}^0 + i(\dot{u}u + \dot{d}d))^2 - (x^1 - i(\dot{u} d + \dot{d}u))^2 \nonumber \]
\[- (\dot{x}^2 + i(\dot{u}u - \dot{d}d))^2\bigg]. \eqno(12) \]
The canonical momenta associated with $e$, $x^0$, $x^1$, $x^2$, $u$ and $d$ are now given by 
\[ \hspace{-3.3cm}p_e = 0 \eqno(13a) \]
\[p_0 = \frac{1}{e} (\dot{x}^0 + i (\dot{u}u + \dot{d}d))\eqno(13b) \]
\[\hspace{.3cm}p_1 = -\frac{1}{e} (\dot{x}^1 + i (\dot{u}d + \dot{d}u))\eqno(13c) \]
\[\hspace{.3cm}p_2 = -\frac{1}{e} (\dot{x}^2 + i (\dot{u}u + \dot{d}d))\eqno(13d) \]
\[\hspace{-1.2cm} \pi_u = i (up_+ - dp_1)\eqno(13e) \]
\[\hspace{-1.2cm} \pi_d = i(dp_- - up_1) \eqno(13f) \]
where $p_{\pm} = p_0 \pm p_2$. Upon defining
\[ \sigma_1 = \pi_u - i (up_+ - dp_1) \eqno(14a) \]
\[ \sigma_2 = \pi_d - i (dp_- - up_1) \eqno(14b) \]
we see that
\[ \left\lbrace \sigma_1, \sigma_1 \right\rbrace = 2ip_+ \eqno(15a) \]
\[ \left\lbrace \sigma_2, \sigma_2 \right\rbrace = 2ip_- \eqno(15b) \]
\[ \left\lbrace \sigma_1, \sigma_2 \right\rbrace = -2ip_1 \eqno(15c) \]
so that
\[ \det  \left\lbrace \sigma_i, \sigma_j \right\rbrace = -4p^2 . \eqno(16) \]
As noted in the preceding section, the secondary constraint of eq. (5) means that the DB cannot be defined for the constraints of eq. (14) when the first class constraints vanish.

One can instead of attempting to define a DB that removes all second class constraints at once, define the DB for a subset of these constraints and proceed in stages as in ref. [19].  In this case, we can first remove the constraint $\sigma_1$, of eq. (14a) by defining 
\[ \left\lbrace A,B \right\rbrace^* = \left\lbrace A,B \right\rbrace - 
\left\lbrace A,\sigma_1 \right\rbrace \frac{1}{2ip_+} \left\lbrace \sigma_1,B \right\rbrace . \eqno(17) \]
(We could have removed the constraint of $\sigma_2$ first by defining
\[ \left\lbrace A,B \right\rbrace^{\#} = \left\lbrace A,B \right\rbrace - 
\left\lbrace A,\sigma_2 \right\rbrace \frac{1}{2ip_-} \left\lbrace \sigma_2,B \right\rbrace .)\eqno(18) \]

With the DB of eq. (17), we find that 
\[\hspace{-1cm} \left\lbrace u, \pi_u \right\rbrace^* = -1/2 \eqno(19a) \]
\[ \left\lbrace u, \pi_d \right\rbrace^* = - p_1/(2p_+) \eqno(19b) \]
\[ \left\lbrace u, u \right\rbrace^* = - 1/(2ip_+) \eqno(19c) \]
\[\hspace{-.2cm}
 \left\lbrace \pi_d, \pi_d \right\rbrace^* =  p_1^2/(2ip_+) \eqno(19d) \]
and so
 \[ \left\lbrace \sigma_2, \sigma_2 \right\rbrace^* = 2ip^2/p_+ .\eqno(20) \]
From eq. (20), we find that by having used a DB to eliminate $\sigma_1$, the constraint $\sigma_2$ has become first class, along with Bosonic first class constraints of eqs. (3a) and (5).  This clarifies how the constraint of eq. (6) is to be decomposed into first and second class constraints when $d = 2+1$.

With the constraints of eqs. (3a, 5, 14b) now all being first class, we can define the generator $G$ of a gauge symmetry transformation by 
\[ G = (a_1 p_e + b\sigma_2 ) + (a_2 p^2) .\eqno(21) \]
We now can make use of eq. (A.13).  Taking the total Hamiltonian to be
\[ H_T = \frac{e}{2}\,p^2 + U_e p_e + U_\sigma \sigma_2 \eqno(22) \]
we have
\[ \left( \frac{Da_1}{Dt}\right) p_e + \left( \frac{Db}{Dt}\right) \sigma_2 + \left( \frac{Da_2}{Dt}\right) p^2 \eqno(23) \]
\[ \hspace{1.5cm}+ \left\lbrace G, H_T \right\rbrace^* - \delta U_ep_e - \delta U_\sigma \sigma_2 = 0 \nonumber \]
which is satisfied by having the coefficient of $p^2$ in this equation vanishing:
\[ \dot{a}_2 - \frac{a_1}{2} - \frac{2i}{p_+} bU_\sigma = 0. \eqno(24) \]
Upon taking $a_2 = a$, we see that together eqs. (21) and (24) lead to 
\[ G = \left( 2\dot{a} - \frac{4i}{p_+} bU_\sigma\right) p_e + b\sigma_2 + ap^2 .\eqno(25) \]

In $G$, $a$ is a Bosonic gauge function and $b$ is a Fermionic one.  $G$ has a somewhat surprising dependence on the (as yet) unrestricted Lagrange multiplier field $U_\sigma$ arising in $H_T$ of eq. (22); this comes about as the DB of the primary constraint $\sigma_2$ with itself is the secondary first class constraint $p^2$ (eq. (20)).

With $G$ given by eq. (25), it is possible to compute the gauge transformation $\delta A = \left\lbrace A,G \right\rbrace^*$ it induces in any dynamical variable $A$.  Since from eq. (17) we have 
\[ \left\lbrace x^0, \pi_d \right\rbrace^* = -\frac{iup_1}{2p_+},\quad \left\lbrace x^0, u\right\rbrace^* = -\frac{u}{2p_+}, \quad 
\left\lbrace x^0, p_0 \right\rbrace^* = 1 \eqno(26a-c) \]
we find that
\[ \delta x^0 = 2ap_0 + ib \left( \frac{up_1}{p_+} + d \right).\eqno(27)\]
Similarly, it follows that
\[ \delta x^1 = 2ap_1 - ib \left( \frac{dp_1}{p_+} + u \right)\eqno(28)\]
\[ \delta x^2 = 2ap_2 + ib \left( \frac{up_1}{p_+} - d \right),\eqno(29)\]
as well as 
\[\delta u = - \frac{bp_1}{p_+} \eqno(30)\]
\[\delta d = -b \eqno(31) \]
\[ \delta e = \left( 2\dot{a} - \frac{4i}{p_+} bU_\sigma \right). \eqno(32) \]
It is now possible to fix $U_\sigma$ by having eqs. (27-32) be transformations that leave the action of eq. (1) invariant.  These equations show that 
\[ \delta Y^0 = 2 \frac{d}{d\tau}(ap_0) - 2i \left( \dot{u}\frac{p_1}{p_+} + \dot{d}\right) b \eqno(33) \]
\[ \delta Y^1 = -2 \frac{d}{d\tau}(ap_1) + 2i \left( \dot{u}+ \dot{d}\frac{p_1}{p_+}\right) b \eqno(34) \]
\[ \delta Y^2 = -2 \frac{d}{d\tau}(ap_2) - 2i \left( \dot{u}\frac{p_1}{p_+} - \dot{d}\right) b \eqno(35) \]
and so 
\[ \delta (Y^\mu Y_\mu) = -4 i\dot{d} b \frac{p_2}{p_+} e  + 2ep^\mu\frac{d}{d\tau} (ap_\mu). \eqno(36) \]

Together, eqs. (13b-d, 32, 36) show that $\delta L = \delta (Y^\mu Y_\mu / (2e))$ is a total derivative provided
\[U_\sigma = -\dot{d} \eqno(37) \]

We now can compare the transformations of eqs. (27-32) with those appearing in the literature, especially the $\kappa$-symmetry transformation of eq. (10).  First of all, the symmetry associated with the parameter $a$ was noted in ref. [2].  We next observe that if we are in $2+1d$, the symmetry of eq. (10) becomes 
\[ \delta x^0 = i \left[ u(-p_1 k + p_- \ell ) + d( - p_+ k + p_1 \ell) \right] \eqno(38a) \]
\[ \delta x^1 = i \left[-d(-p_1 k + p_- \ell ) - u( - p_+ k + p_1 \ell) \right] \eqno(38b) \]
\[ \delta x^2 = i \left[ u(-p_1 k + p_- \ell ) - d( - p_+ k + p_1 \ell) \right] \eqno(38c) \]
\[ \delta u = -p_1 k + p_- \ell \eqno(38d) \]
\[ \delta d = -p_+ k + p_1 \ell \eqno(38e) \]
\[ \delta e = 4ie(\dot{d} k - \dot{u} \ell) \eqno(38f)\]
upon identifying $\kappa^T = (k, \ell) = (k^*, \ell^*)$.  If we now make use of eq. (36) and set $k = b/p_+$, $\ell = 0$ we find that eqs. (38) are the same as the $b$ dependent part of eqs. (27-32).  This shows that the $\kappa$-symmetry is in part (the $k$-dependent part) generated by the first class constraints ($p_e,\, p^2,\, \sigma_2$) in the model once the constraint $\sigma_1$ has been eliminated by using the DB of eq. (17).  To obtain the other ($\ell$-dependent) part of the $\kappa$-symmetry, one need only reverse the roles of $\sigma_1$ and $\sigma_2$.  Upon defining the DB $\left\lbrace A,B \right\rbrace^{\#}$ of eq. (18) to eliminate $\sigma_2$, one find that $\sigma_1$ becomes a first class constraint; the first class constraints $(p_e,\, p^2,\, \sigma_1)$ then can be used to find a generator for the $\ell$-dependent parts of the transformations of eq. (38).

We now will examine the DB algebra of the generator $G$ of eq. (25) with $U_\sigma$ being given by eq. (37).  If we define $G_i$ to be the generator associated with gauge parameters $a_i$ and $b_i$, then in computing 
$\left\lbrace G_i, G_j \right\rbrace^*$ we are confronted with an ill defined DB $\left\lbrace \frac{4i}{p_+} b_1\dot{d} p_e, \; b_2\pi_d\right\rbrace^*$; if this is replaced by $\left\lbrace - \frac{d}{d\tau} \left(\frac{4ib_1 p_e}{p_+}\right) d, \; b_2\pi_d\right\rbrace^*$ then it follows that 
\[\left\lbrace G_i, G_j \right\rbrace^* = \left(\frac{-2ib_1b_2}{p_+}\right) p^2 - 4i\bigg[ \left( \frac{d}{d\tau} (b_1b_2)\right)\left( \frac{p_e}{p_+}\right)\nonumber \]
\[+ 2b_1b_2 \frac{d}{d\tau} \left(\frac{p_e}{p_+}\right) \bigg]. \eqno(39) \]
We thus see that the DB of two generators $G_1$ and $G_2$ is a purely Bosonic generator with
\[ a = \frac{-2ib_1b_2}{p_+} \eqno(40) \]
provided we use the equations of motion $\dot{p}_e = \dot{p}_+ = 0$.

We now turn to an examination of the superparticle in $d = 3+1$ dimensions.  Most results of this section can be generalized to this higher dimension.

\section{d = 3+1}

The spinor $\theta$ appearing in eq. (1) now takes the form given in eq. (A.6), 
\[ \theta = (T, \quad B, \quad iB^*, \quad -iT^*)^T\eqno(41)\]
provided $\theta$ is Majorana.  The quantities $T$ and $B$ are complex Fermionic functions of $\tau$. The quantities $Y^\mu$ of eq. (2) are thus given by 
\[ Y^0 = x^0 + i(\dot{T}^* T + \dot{B}^* B + \dot{T} T^* + \dot{B}B^*)\eqno(42a) \]
\[ Y^1 = x^1 + i(\dot{B}^* T + \dot{T}^* B + \dot{T} B^* + \dot{B}T^*)\eqno(42b) \]
\[ Y^2 = x^2 + i(i\dot{B}^* T - i \dot{T}^* B + i \dot{T} B^* - i \dot{B}T^*)\eqno(42c) \]
\[ Y^3 = x^3 + i(\dot{T}^* T - \dot{B}^* B + \dot{T} T^* - \dot{B}B^*) .\eqno(42d) \]
Eqs. (3a) and (3b) again hold; the momenta conjugate to $(T,\;B,\; T^*\;B^*)$ respectively lead to the primary constraints
\[ \lambda_1 = \pi_T -i(T^*\; p_+ + B^*\; p_I ) \eqno(43a) \]
\[ \lambda_2 = \pi_B -i(B^*\; p_- + T^*\; p_I^* ) \eqno(43b) \]
\[ \lambda_3 = \pi_T^* -i(T\; p_+ + B\; p_I^* ) \eqno(43c) \]
\[ \lambda_4 = \pi_B^* -i(B \; p_- + T\; p_I ) \eqno(43d) \]
where $p_\pm = p_0 \pm p_3$ and $p_I = p_1 + ip_2$.

Just as in the $3d$ case, the constraint $\sigma_1$ of eq. (14a) can be eliminated by the DB of eq. (17), so also the constraints $\lambda_1$ and $\lambda_3$ of eqs. (43a,c) can be eliminated by defining the DB
\[ \left\lbrace A,B \right\rbrace^* = \left\lbrace A, B \right\rbrace - \frac{1}{2ip_+} \left[ \left\lbrace A, \lambda_1 \right\rbrace \left\lbrace \lambda_3, B \right\rbrace + \left\lbrace A, \lambda_3 \right\rbrace \left\lbrace \lambda_1, B \right\rbrace \right]. \eqno(44) \]
From eq. (44) it follows that
\[ \left\lbrace \lambda_2, \lambda_2 \right\rbrace^* = 0 = \left\lbrace \lambda_4, \lambda_4 \right\rbrace^* \eqno(45a) \]
\[\left\lbrace \lambda_2, \lambda_4 \right\rbrace^* = 2ip^2/p_+ \eqno(44b) \]
much as with eq. (20).

With the total Hamiltonian being given by 
\[ H_T = \frac{ep^2}{2} + U_ep_e + U_2\lambda_2 + U_4\lambda_4 \eqno(46) \]
we can show from the HTZ eq. (A.13)
\[ G = \left[ 2\dot{a} - \frac{4i}{p_+} (b_2 U_4 + b_4U_2)\right]p_e + b_2 \lambda_2 + b_4 \lambda_4 + ap^2 \eqno(47) \]
(much as eq. (25)) where $a$ is a Bosonic gauge parameter and $b_2, b_4$ are Fermionic gauge parameters.

If we let the gauge function $\kappa$ in eq. (10) be given by 
\[ \kappa = (M,\;N,\; iN^*,\; -iM^*)^T \eqno(48) \]
while in $(3+1)d$, then it can be shown that the transformations generated by $G$ in eq. (47) (ie, $\delta A = \left\lbrace A, G \right\rbrace^*$) are those of eq. (10) provided
\[ M = b_2/p_+ = b_4^*/p_+,\quad N = 0, \eqno(49) \]
with
\[ U_2 = -\dot{B}^*, \quad U_4 = - \dot{B} . \eqno(50) \]
Furthermore, if one were to eliminate the primary constraints $\lambda_2$ and $\lambda_4$ through use of the DB
\[ \left\lbrace A,B \right\rbrace^{\#} =  \left\lbrace A,B \right\rbrace - \frac{1}{2ip_-} \left[ \left\lbrace A,\lambda_2 \right\rbrace \left\lbrace  \lambda_4, B \right\rbrace + \left\lbrace A,\lambda_4 \right\rbrace \left\lbrace  \lambda_2, B \right\rbrace \right] \eqno(51) \]
then $(\lambda_1,\; \lambda_3)$ become first class constraints; together $(p_e,\; p^2, \;\lambda_1, \; \lambda_3)$ can then be used to find a generator $G$ which generates the $N$ dependent part of the $\kappa$-transformation eq. (10) in $(3+1)d$.  This is in analogy with what was shown to occur in $(2+1)d$ in the preceding section. 

\section{Discussion}

We have demonstrated that the primary constraints associated with the Fermionic canonical momenta in $(2+1)d$ and $(3+1)d$ can be segregated into two parts; if a DB is used to eliminate one of these parts, then the remaining part becomes first class.  Subsequently, the first class constraints can be used to find a generator of one portion of the $\kappa$-symmetry through use of the HTZ procedure.  The other portion of the $\kappa$-symmetry can be found by reversing the roles of the two parts of the Fermionic primary constraints. 

Although this analysis is not manifestly convariant, it does shed some light on the nature of the constraints arising when considering the superparticle action and establishes how these constraints are related to the $\kappa$-symmetry.  It is hoped that it will be possible to consider the superparticle in higher dimensions and also the superstring using this approach and that some insight into quantization of this model can be gained. 

\section*{Acknowledgements}

Roger Macleod stimulated this research.

\newpage
\section*{Appendix}

We use the following conventions:
\subsection*{ I - Spinors in $2 + 1d$}

We use the flat metric $\eta_{\mu\nu} = \textrm{diag} (+ - -)$ with Dirac matrices (in terms of the Pauli matrices $\sigma^i$)
\[ \gamma^0 = \sigma^2 \qquad \gamma^1 = i\sigma^3 \qquad \gamma^2 = i\sigma^1  .\eqno(A.1) \]
These satisfy the conditions
\[ \gamma^\mu\gamma^\nu = \eta^{\mu\nu} + i\epsilon^{\mu\nu\lambda} \gamma_\lambda \eqno(A.2) \]
and
\[ \gamma^0 \gamma^\mu\gamma^0 = -\gamma ^{\mu T} =  \gamma^{\mu\dagger}. \eqno(A.3) \]
Spinors are taken to be Majorana so that $\psi = C\overline{\psi}^T$ where 
\[ \overline{\psi} = \psi^\dagger \gamma^0, \qquad C = -\gamma^0 \eqno(A.4) \]
so that $\psi = \psi^* = (u, d)^T$ with $u$ and $d$ real. 

\subsection*{II - Spinors in $3 + 1d$}

The Dirac matrices are now given by 
\[ \gamma^0 = \left( \begin{array}{cc}
0 & 1 \\
1 & 0 \end{array}
\right)\qquad
\vec{\gamma} = \left( \begin{array}{cc}
0 & -\vec{\sigma} \\
\vec{\sigma} & 0 \end{array}
\right)
\eqno(A.5) \]
and $C \equiv \gamma^0\gamma^2$ so that if $\psi = C \overline{\psi}^T (\overline{\psi} \equiv \psi^\dagger \gamma^0)$ then
\[ \psi = \left( \begin{array}{c}
T \\
B\\
iB^*\\
-iT^* \end{array}
\right) \quad \textrm{and} \quad
\overline{\psi} = (-iB,\, iT, \, T^*, \, B^*). \eqno(A.6) \]

\subsection*{III - Canonical formalism}

If $L = L(q_i, \psi_i;\; \dot{q}_i, \dot{\psi}_i)$ is a Lagrangian dependent on Bosonic variable $q_i$ and Fermionic variables $\psi_i$, then we define the canonical momenta
\[ p_i = \frac{\partial L}{\partial \dot{q}_i}\qquad
\pi_i = \frac{\partial L}{\partial \dot{\psi}_i}
\eqno(A.7) \]
and the canonical Hamiltonian
\[ H = \dot{q}_i p_i + \dot{\psi}_i \pi_i - L \eqno(A.8) \]
where we use the ``left derivative'' for Fermionic variables $\theta^A$
\[ \frac{d}{d\theta^A} (\theta^B \theta^C) = \delta^{AB}\theta^C - \delta^{AC} \theta^B \eqno(A.9) \]
\[ \frac{d}{dt} F(\theta (t)) = \dot{\theta}(t) F^\prime (\theta(t)).\nonumber\]
Note also that $(\theta^A\theta^B)^\dagger = \theta^{B\dagger} \theta^{A\dagger}$. \\ 
For Bosonic variables $B_i$ and Fermionic variables $F_i$ we employ the Poisson Brackets (PB) 
\[ \left\lbrace B_1, B_2 \right\rbrace = (B_{1,q} B_{2,p} - B_{2,q} B_{1,p}) + (B_{1,\psi} B_{2,\pi} - B_{2,\psi} B_{1,\pi})\nonumber \]
\[ \left\lbrace B, F \right\rbrace = (B_{,q} F_{,p} - F_{,q} B_{,p}) + (B_{,\psi} F_{,\pi} + F_{,\psi} B_{,\pi}) = - \left\lbrace F, B \right\rbrace\nonumber \]
\[ \left\lbrace F_1, F_2 \right\rbrace = (F_{1,q} F_{2,p} + F_{2,q} F_{1,p}) - (F_{1,\psi} F_{2,\pi} + F_{2,\psi} F_{1,\pi})\eqno(A.10) \]

\subsection*{IV - The Gauge Generator}

If one has a set of first class constraints $\gamma_{a_{i}}$ ($i$-generation of the constraint) and a canonical Hamiltonian $H_c$, then the first step is to eliminate the second class constraints $\theta_{a_{i}}$ through use of Dirac Brackets $\left\lbrace , \right\rbrace^*$ in place of Poisson Brackets $\left\lbrace , \right\rbrace$ .  One then forms the extended action 
\[ S_E = \int d\tau \left[ \dot{q}_i p_i + \dot{\psi}_i\pi_i - H_c - U_{a_{i}} \gamma_{a_{i}} \right] \eqno(A.11) \]
where $U_{a_{i}}$ are a set of Lagrange multipliers.  The total action $S_T$ is found by setting $U_{a_{i}} = 0$ for $i \geq 2$ in eq. (A.11); it is equivalent to the classical action $S_C = \int d\tau \, L$.  In the HTZ approach [11] one find that a generator of gauge transformations $G$ (ie, $\delta A = \left\lbrace A,G \right\rbrace^*$) given by 
\[ G = \mu_{a_{i}} \gamma_{a_{i}} \eqno(A.12) \]
leaves $S_E$ invariant provided 
\[ 0 = \int d\tau \left[ \frac{D\mu_{a_{i}}}{Dt} \gamma_{a_{i}} + \left\lbrace G, H_c + U_{a_{i}} \gamma_{a_{i}} \right\rbrace^* - \delta U_{a_{i}} \gamma_{a_{i}} \right]. \eqno(A.13) \]

The generator $G$ of a gauge transformation that leaves $S_T$ (and hence $S_C$) invariant can be found by setting $U_{a_{i}} = \delta U_{a_{i}} = 0$ ($i \geq 2$) in eq. (A.13) and systematically solving for $\mu_{a_{i}}$.  ($D/Dt$ denotes the total time derivative exclusive of dependence on the canonical position and momentum.)

\end{document}